\documentclass[aps,twocolumn,showpacs,amsmath,amssymb,pra,superscriptaddress,floatfix,longbibliography]{revtex4-2}
\usepackage{
  amsmath,
  booktabs,
  braket,
  esdiff,
  graphicx,
  hyperref,
  nicefrac,
  siunitx,
  xcolor,
  xspace,
  caption,
  subcaption
}
\usepackage{lipsum}
\hypersetup{colorlinks,urlcolor=blue ,citecolor=blue ,linkcolor=blue}
\def\unit #1 #2 {\SI{#1}{#2}\xspace}
\usepackage{xcolor}
\DeclareSIUnit\gauss{G}
\DeclareCaptionLabelSeparator{dot}{. }
\makeatletter
\def\justified{
	\let\\\@normalcr
	\@rightskip\z@skip \rightskip\@rightskip
	\leftskip\z@skip
	\parindent 0em\relax
	\setlength{\parfillskip}{0pt plus 1fil}}
\DeclareCaptionJustification{justified}{\justified}
\newcommand{\Er}{\ensuremath{^{166}}{\rm Er} }

\newcommand{\as}{\ensuremath{a_{\rm s}}}

\newcommand{\tht}{\ensuremath{t_{\rm h}}}
\newcommand{\Erec}{\ensuremath{E_{\rm rec}}}
\newcommand{\meank}{\ensuremath{\langle\left| q_z \right|\rangle}\,}

\begin{document}
		\title{Strongly dipolar gases in a one-dimensional lattice: Bloch oscillations and matter-wave localization}
	\author{G.\,Natale}
	\affiliation{Institut f\"ur Quantenoptik und Quanteninformation, \"Osterreichische Akademie der Wissenschaften, Technikerstra{\ss}e 21a, 6020 Innsbruck, Austria}	
	\affiliation{Institut f\"ur Experimentalphysik, Universit\"at Innsbruck, Technikerstra{\ss}e 25, 6020 Innsbruck, Austria}
	\author{T.\,Bland}
	\affiliation{Institut f\"ur Experimentalphysik, Universit\"at Innsbruck, Technikerstra{\ss}e 25, 6020 Innsbruck, Austria}
	\author{S.\,Gschwendtner}
	\affiliation{Institut f\"ur Experimentalphysik, Universit\"at Innsbruck, Technikerstra{\ss}e 25, 6020 Innsbruck, Austria}
	\author{L.\,Lafforgue}
	\affiliation{Institut f\"ur Experimentalphysik, Universit\"at Innsbruck, Technikerstra{\ss}e 25, 6020 Innsbruck, Austria}
	\author{D.\,S.\,Gr\"un}
	\affiliation{Institut f\"ur Experimentalphysik, Universit\"at Innsbruck, Technikerstra{\ss}e 25, 6020 Innsbruck, Austria}
	\author{A.\,Patscheider}
	\affiliation{Institut f\"ur Experimentalphysik, Universit\"at Innsbruck, Technikerstra{\ss}e 25, 6020 Innsbruck, Austria}
	\author{M.\,J.\,Mark}
	\affiliation{Institut f\"ur Quantenoptik und Quanteninformation, \"Osterreichische Akademie der Wissenschaften, Technikerstra{\ss}e 21a, 6020 Innsbruck, Austria}
	\affiliation{Institut f\"ur Experimentalphysik, Universit\"at Innsbruck, Technikerstra{\ss}e 25, 6020 Innsbruck, Austria}
	\author{F.\,Ferlaino}
	\affiliation{Institut f\"ur Quantenoptik und Quanteninformation, \"Osterreichische Akademie der Wissenschaften, Technikerstra{\ss}e 21a, 6020 Innsbruck, Austria}
	\affiliation{Institut f\"ur Experimentalphysik, Universit\"at Innsbruck, Technikerstra{\ss}e 25, 6020 Innsbruck, Austria}

\date{\today}
	
\begin{abstract}
Three-dimensional quantum gases of strongly dipolar atoms can undergo a crossover from a dilute gas to a dense macrodroplet, stabilized by quantum fluctuations.
Adding a one-dimensional optical lattice creates a platform where quantum fluctuations are still unexplored, and a rich variety of new phases may be observable. We employ Bloch oscillations as an interferometric tool to assess the role quantum fluctuations play in an array of quasi-two-dimensional Bose-Einstein condensates.
Long-lived oscillations are observed when the chemical potential is balanced between sites, in a region where a macrodroplet is extended over several lattice sites.
Further, we observe a transition to a state that is localized to a single lattice plane--driven purely by interactions--marked by the disappearance of the interference pattern in the momentum distribution.
To describe our observations, we develop a discrete one-dimensional extended Gross-Pitaevskii theory, including quantum fluctuations and a variational approach for the on-site wavefunction. This model is in quantitative agreement with the experiment, revealing the existence of single and multisite macrodroplets, and signatures of a two-dimensional bright soliton.
\end{abstract}

\maketitle
The dipole-dipole interaction (DDI) between magnetic atoms in an ultracold quantum gas has been key to the discovery of supersolids \cite{Tanzi2019ooa,Boettcher2019tsp,Chomaz2019lla} and macrodroplets \cite{Chomaz2016qfd,schmitt2016sbd}, new states of matter with extremely intriguing and counter-intuitive properties \cite{Norcia2021nof,chomaz2022dpa}. Macrodroplets are macroscopic quantum states that behave in many ways like liquid droplets \cite{schmitt2016sbd,Chomaz2016qfd,Cabrera2018qld,Semeghini2018sbq}. They are at least an order of magnitude denser than normal Bose-Einstein condensates (BECs), and can be self-bound. They exist in a parameter regime in which mean-field theories predict the collapse of the entire system when the attractive dipolar interactions overcome the repulsive contact interactions. Instead, the system remains surprisingly stable thanks to the so-called quantum fluctuations, thus providing one of the rare examples where beyond-mean-field interactions substantially change the ground state of the system \cite{Petrov2015qms,Baillie2016sbd}. 
Although the functional form of the beyond-mean-field term, otherwise known as the Lee-Huang-Yang (LHY) correction \cite{Lima2011qfi}, is still subject to intense study and debate \cite{cikojevic2019universality,ota2020beyond}, its importance is now undoubted. Isolating beyond-mean-field effects may be crucial to settle disputes on its validity, particularly in dipole dominated systems; however, it is very difficult to have access to individual interaction contributions. Though, the differing atom number scaling between mean-field and LHY contributions provide a promising method to differentiate between them.

Optical lattices enable powerful interferometric approaches to, e.g., measure with high precision the zero-crossing of the scattering length or of the mean-field interaction with the so-called Bloch oscillation (BO) technique \cite{Roati2004aiw,Ferrari2006llb,Gustavsson2008coi,Fattori2008mdi,Fattori2008mdi}, and to achieve an accurate determination of the background scattering length via lattice spectroscopy in Hubbard models \cite{Chomaz2016qfd,Baier2018roa,patscheider2021accurate}. Moreover, the presence of the lattice itself may change completely the phase diagram of the system, as shown in seminal experiments with contact interacting gases \cite{Bloch2008mbp,Bloch2012qsw,Gross2017qsw}. Unique phenomena are predicted with the addition of long-range DDIs \cite{Trefzger2011udg,Dutta2015nsh}.
Experiments with lattice-confined atomic dipolar gases have already shown important results, e.g., the realization of extended Bose-Hubbard models \cite{Baier2016ebh} and spin models \cite{dePaz2013nqm,Lepoutre2019ooe,Gabardos2020rot,Patscheider2020cde} in three-dimensional (3D) lattices. In 2D lattices, forming quasi-1D tubes, suppression of dipolar relaxation \cite{Pasquiou2011sra} and the controlled breakdown of integrability \cite{Tang2018tni} have been observed.
Instead, up to now, 1D lattices, forming an array of quasi-2D layers, have been used with large wavelengths to load a single pancake trap \cite{Koch2008soa}, or multi-layer traps to study the role of DDI in the stability against collapse \cite{Muller2011soa}.
Further, theoretical proposals have suggested that the DDI between layers not only can lead to modifications within each layer \cite{Koberle2009,Klawunn2009hme,Wilson2011esi,Rosenkranz2013} but also to inter-layer bound states \cite{Wang2007qpt,Trefzger2009psp,Macia2014spv}. Other works predict the existence of bright-soliton structures along the lattice \cite{Gligoric2008bsi} or anisotropic on-site solitons \cite{Tikhonenkov2008asi,Raghunandan2015tdb}. However, those proposals lack the important stabilization mechanism
given by the LHY term, which is known to provide many new phases in continuous systems (e.g.\,harmonically trapped), opening up many questions: What is the ground state of an attractive dipolar gas in a 1D lattice potential? Can droplets be delocalized over many lattice planes? Will solitonic solutions continue to exist?

In the present work, we study an erbium dipolar gas in a 1D optical lattice with dominantly attractive DDI.
We employ BOs as an interferometric tool to probe the interaction contributions of the system, and to isolate the role of beyond-mean-field effects.
We find long-lived oscillations, associated with a minimum in the dephasing rate, close to the cancellation point between mean-field and beyond-mean-field interactions, and at scattering lengths significantly shifted from the expected mean-field result.
We develop a discrete effective 1D  extended Gross-Pitaevskii equation (eGPE) with variational transverse widths \cite{Blakie2020vtf,Blakie2020sia}. We find that this minimum occurs when the chemical potentials on each site are equal, not the energies--as has been employed successfully in contact interaction dominated systems \cite{Gustavsson2008coi,Fattori2008mdi}--due to the difference in density scaling between the interactions. The close correspondence between theory and experiment shows the validity of the LHY prediction, even while highly inhomogeneous densities are expected to break the local density approximation \cite{Lima2011qfi}.
Moreover, we see that for low scattering lengths the system undergoes a structural transition to a single localized 2D plane, signifying an important new way to generate systems in reduced geometries through varying the interactions alone.
Finally, using our theoretical model we produce a full phase diagram of the system, revealing the impact of the LHY contribution to the predicted 2D anisotropic soliton state \cite{Tikhonenkov2008asi}, which is instead morphed into a droplet solution at high atom numbers. Though, promisingly, we still find soliton-like solutions exist.

\begin{figure}[t!]
\centering
\includegraphics[width=\columnwidth]{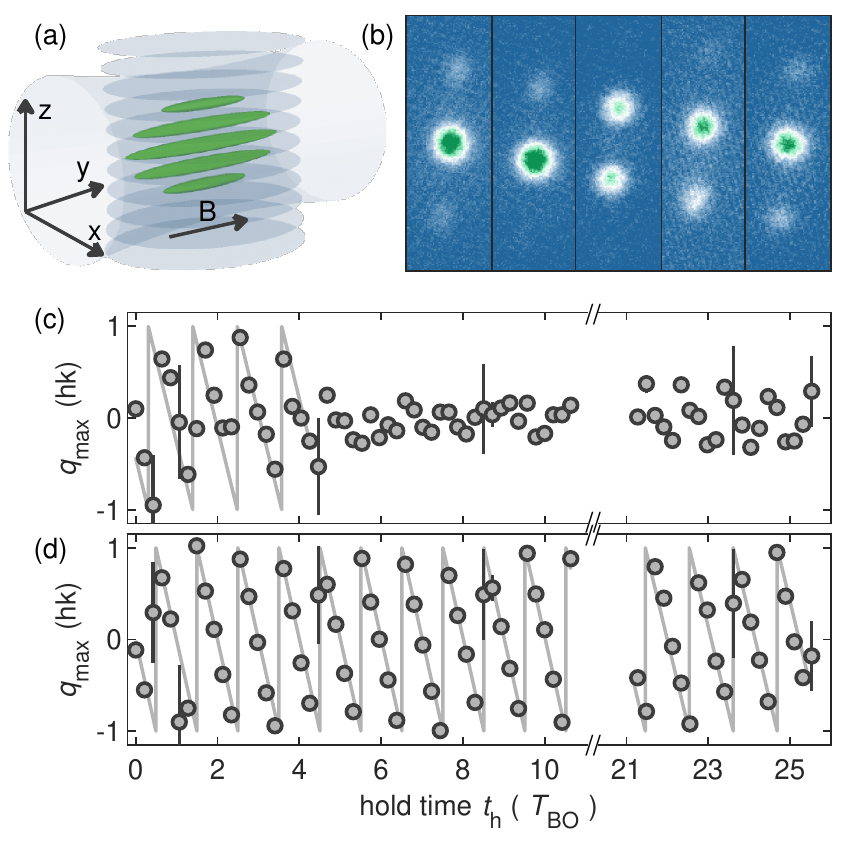} %0.75\textwidth
\caption{\label{Fig1}
\textbf{Bloch oscillations of a dipolar BEC in a one-dimensional optical lattice.} \textbf{(a)} Sketch of our experiment, consisting of a 1D optical lattice in the $z$-direction, loaded with an erbium BEC from an optical dipole trap with trapping frequencies $\omega_{x,y,z} = 2\pi\times(240(3),30(3),217(1))$ \SI{}{\hertz}. Gravity acts along $z$. \textbf{(b)} Absorption images after TOF showing the momentum distributions during one Bloch cycle.
\textbf{(c,d)} Evolution of the peak position of the momentum distribution for $\as$ = (71.6(1.0), 59.8(1.0))$\,\SI{}{\bohr}$,  respectively. A sawtooth fit (solid grey) to the data yields $T_{\textrm{BO}}=\SI{0.469(4)}{\milli\second}$, consistent with the expected value $T_{\textrm{BO}}=2k/(mg_\text{grav})$.
The error bars represent the standard error on the mean over 4-6 repetitions.}
\end{figure}

In the experiment, we prepare a degenerate dipolar gas of erbium atoms in a one-dimensional optical lattice as follows. We start with a dipolar quantum gas of $5 \times 10^4$ spin-polarized $\Er$ atoms confined in a cigar-shaped optical dipole trap \cite{Aikawa2012} elongated along $y$. Typical BEC fractions range from 60\% to 80\%. The dipolar length for $\Er$ is fixed at $a_\text{dd}=\SI{66.5}{\bohr}$, where $\SI{}{\bohr}$ is the Bohr radius. We tune the contact interaction between atoms and therefore the $s$-wave scattering length, $\as$, via Feshbach resonances \cite{Chin2010fri, Frisch2014qci, Chomaz2016qfd,  suppmat} by changing the absolute value of a bias magnetic field $|\mathbf{B}|$. We fix the orientation of $\mathbf{B}$ to be along the weak axis ($y$) of the trap, making the DDI dominantly attractive \cite{Chomaz2016qfd,chomaz2022dpa}.

Once the harmonically-trapped cloud is prepared at the desired $\as$, we switch on a 1D optical lattice, aligned along the gravity direction ($z$); see Fig.\,\ref{Fig1}(a). The vertical lattice is created by retro-reflecting a $\lambda=\SI{1064}{\nano\meter}$ laser beam. We load the planes by exponentially increasing the lattice depth $V_0$ to $8\,\Erec$ in $\SI{20}{\milli\second}$, where $\Erec= \hbar^2k^2/2m=h \times \SI{10.5}{\kilo\hertz}$. Here, $\hbar=h/2\pi$ is the reduced Planck's constant ($h$), $m$ is the mass of $\Er$ atoms and $k = 2\pi/\lambda$ is the wave-vector of the lattice. The 1D lattice forms an array of tightly confined quasi-2D planes with a trap frequency along the tight direction $\omega_z \simeq 2\pi \times \SI{6}{\kilo\hertz}$, corresponding to an harmonic oscillator length $z_\text{ho}= \SI{100}{\nano\meter}$. The tunnelling rate, $J$, between planes is about $h \times \SI{33}{\hertz}$. For these 1D lattice parameters, $\hbar \omega_z > k_BT$ and the system is kinematically 2D \cite{petrov2001ici}.

We first aim at inducing Bloch oscillations to interferometrically assess the role of beyond-mean-field effects and test the validity of the LHY term. We thus suddenly switch off the dipole trap and let the system evolve in the combined lattice and gravitational potential for a variable hold time $\tht$. Finally, using standard absorption imaging after $\SI{30}{\milli\second}$ of time-of-flight (TOF), we record the evolution of the momentum distribution and extract the position of the main peak, $q_{\textrm{max}}$, as a function of $\tht$. Figure\,\ref{Fig1}(b) shows an exemplary set of absorption images during a single Bloch period $T_{\textrm{BO}}$. We observe the key paradigm of BOs, i.e.\,the linear increase of the mean momentum due to the acceleration and the Bragg reflection occurring at the border of the Brillouin zone \cite{Ben1996boo}, well described by fitting a sawtooth function to $q_{\textrm{max}}$.

\begin{figure}[t]
\centering
\includegraphics[width=\columnwidth]{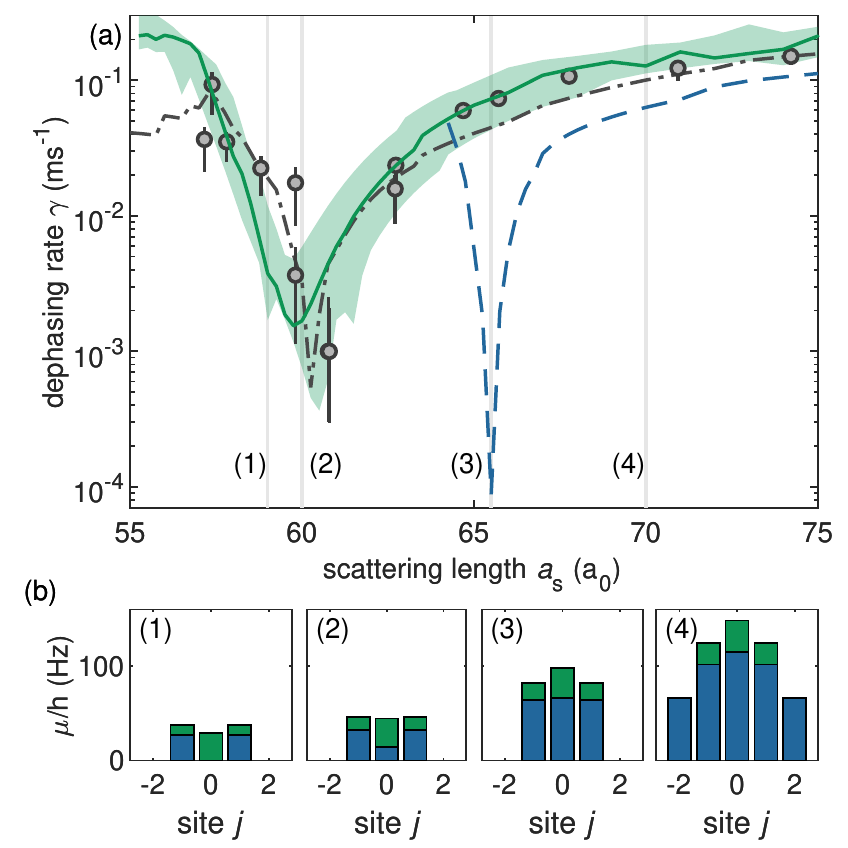} 
\caption{\label{Fig2}\textbf{Dephasing rate and chemical potential distributions.} \textbf{(a)}\,Experimental dephasing rate $\gamma$ (circles) as a function of scattering length $\as$. The green solid line shows the theory result, with an uncertainty region (shaded area) accounting for 20\% atom number variation. The blue dashed line shows the theory expectation without LHY. The gray dot-dashed line gives the prediction of the semi-analytic approximation for $\gamma$. Error bars show the 68\% confidence interval \cite{suppmat}. \textbf{(b)}\,Chemical potential per lattice site $\mu_j$ extracted from the discrete model for $\as=(59,\,60,\,65.5,\,70)\,\SI{}{\bohr}$ $(1,\,2,\,3,\,4)$. The green area depicts the LHY contribution to $\mu_j$.}
\end{figure}

The high sensitivity of BOs to interactions \cite{Gustavsson2008coi,Fattori2008aiw} clearly appears by tracing the evolution for two different $\as$ (see Fig.\,\ref{Fig1}(c,d)), as the interaction dependence is encoded into the dephasing rate. For a contact-dominated gas ($a_\text{dd}<\as=\SI{90}{\bohr}$, Fig.\,\ref{Fig1}(c)), we see that the BOs vanish within a few $T_{\textrm{BO}}$. On the contrary, decreasing $\as$, and thereby going into the regime where contact interactions and DDI nearly compensate each other ($\as=\SI{60}{\bohr}$, Fig.\,\ref{Fig1}(d)), we observe persisting oscillations for more than $25$ Bloch cycles, set by our limited observation time \cite{suppmat}. To systematically study this effect, we repeat the BO measurements for different values of $\as$, and extract the corresponding dephasing rate $\gamma$ \cite{suppmat}. As shown in Fig.\,\ref{Fig2}(a), we observe a resonant-type behavior with $\gamma$ showing a pronounced dip with a minimum at $\as=\SI{61}{\bohr}$. This minimum is clearly different to the point $\as \approx a_\text{dd}$, where the variance of the mean-field energies across different lattice sites cancel \cite{Fattori2008mdi}, which would be expected from previous observations \cite{Gustavsson2008coi,Fattori2008aiw}.

To get further insight on the origin of the minimum, we develop a discrete effective 1D eGPE, 
inspired by the close correspondence between predictions from discrete models and experimental observations in non-dipolar \cite{fallani2008bec,Morsch2006dob} and weakly dipolar \cite{Fattori2008mdi} BECs.
We separate the 3D wavefunction into radial and axial contributions, allowing for a variational anisotropic radial width and thus maintaining the 3D character \cite{Blakie2020vtf}. Along the lattice direction ($z$), we further decompose the wavefunction, $\psi(z, t)$, as a sum of Wannier functions $w(z)$ of the lowest energy band over all lattice sites: $\psi(z, t) = \sqrt{N} \sum_j c_j(t) w(z - z_j)$, where $N$ is the atom number and $c_j(t)$ the complex wavefunction amplitude on lattice site $j$, leading to a set of discrete effective 1D eGPEs, each including mean-field and beyond mean-field interactions. For the beyond-mean-field interaction, the 3D form of the LHY still fully applies since  the contact interaction energy exceed the confinement energy scale  \cite{zin2021quantum,suppmat}. However, our system may also open to further studies on the 2D to 3D crossover of the LHY. We solve these equations coupled to a minimization of the energy functional with respect to the variational parameters to determine the ground states, benchmarking them against the full 3D theory. We then perform dynamic simulations of the expected time evolution \cite{suppmat}, giving an accurate dephasing rate (solid line) in Fig.\,\ref{Fig2}(a) without free parameters.

In previous studies, the point of minimum dephasing was found to occur when the mean-field interaction energies vanish or cancel. We isolate the mean-field contribution by removing beyond-mean-field effects from our simulations (dashed line in Fig.\,\ref{Fig2}(a)), predicting a minimum at $\as \approx a_\text{dd}$. However, this is in clear contradiction with our experimental observations by a shift of 6a$_0$ and a different overall shape due to the different scaling of the LHY term with the density. Without LHY, the cancellation of mean-field energies, $E^j_\text{MF}$, is equivalent to the cancellation of onsite chemical potentials, given by $\mu_j = 2E^j_\text{MF}/|c_j|^2$. Note, $\mu_j$ dictates the wavefunction phase winding on each site through $c_j =|c_j|e^{-i\mu_j t/\hbar}$. Reintroducing quantum fluctuations, we obtain $\mu_j = (2 E^j_\text{MF} + 5/2 E^j_\text{BMF})/|c_j|^2$, where the 5/2 appears due to the $|c_j|^5$ density scaling in the beyond-mean-field energy $(E^j_\text{BMF})$. Figure\,\ref{Fig2}(b) shows $\mu_j$ from the ground state calculation for four scattering lengths, additionally indicating the contribution of the LHY correction.

We observe that the point of minimal dephasing in the experiment is close to the point where the variance of $\mu_j$ is minimized \footnote{The total (MF + LHY) interaction energies cancel at $62a_0$.}. Indeed, within a semi-analytic approximation (see Ref.\,\cite{suppmat} for details), we find a direct relationship between $\gamma$ and $\mu_j$, which reads $\gamma\propto|\mu_1-\mu_{0}|$ when 3 lattice sites ($j=-1,0,1$) are occupied. This model can be extended to 5 lattice sites, giving the dot-dashed line (Fig.\,\ref{Fig2}(a)) which reproduces very well the system behaviour \cite{suppmat}. Interestingly, measuring the dephasing rate through the chemical potential is ubiquitous to systems with arbitrary interaction potentials.

\begin{figure}[t]
\centering
\includegraphics[width=\columnwidth]{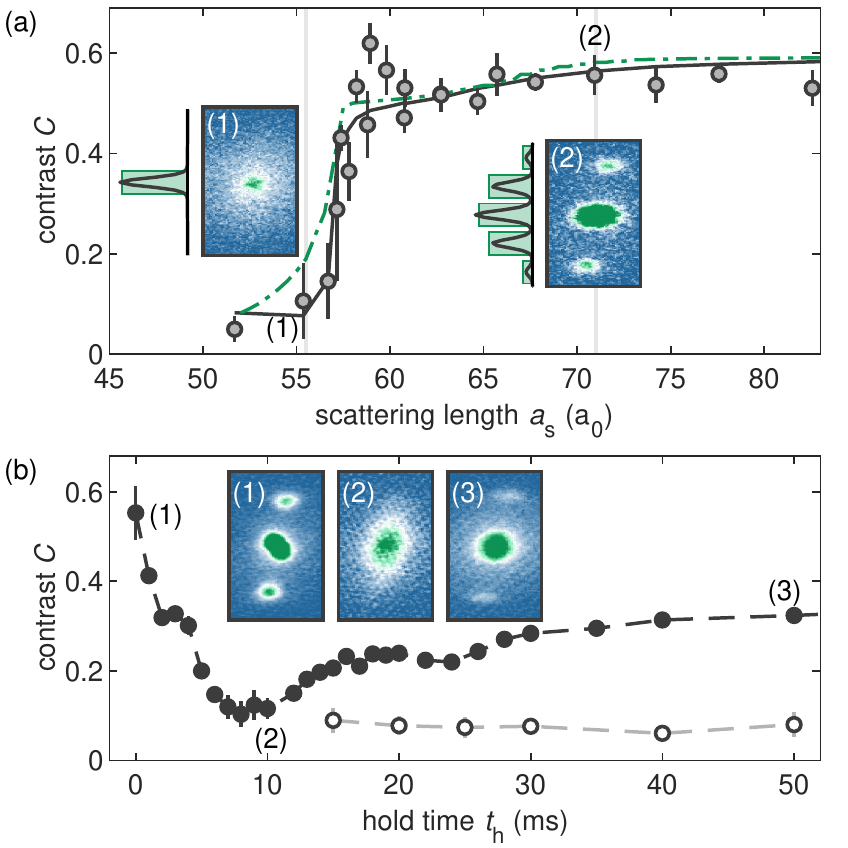} %0.75\textwidth
\caption{\label{Fig3}
\textbf{Interaction-induced localization.} \textbf{(a)}\,Contrast of the interference pattern after loading the lattice at different $\as$. The green dot-dashed (black solid) line represents the result of the 1D discrete model (3D eGPE) multiplied by 0.7. The insets show the respective density distributions along $z$ of the 1D discrete model (bars) and 3D eGPE (lines) and corresponding experimental averaged interference patterns after TOF expansion (1,2). \textbf{(b)}\,Dynamic evolution of the contrast quenching back  (filled circles) or holding $\as$ (open circles); see text. The error bars represent the standard error on the mean over 4-6 repetitions.}
\end{figure}

Surprisingly, by further decreasing the scattering length below $\SI{57}{\bohr}$, no BOs nor interference peaks are visible anymore. We observe at the initial instant ($t_h=\SI{0}{\milli\second}$) that the momentum distribution is already spread over the entire first Brillouin zone.
To quantify this, we study the contrast, $C$, of the interference pattern of the initial momentum distribution as a function of $\as$, see Figure \ref{Fig3}(a).
We extract $C$, defined as the amplitude of the momentum peaks at $\pm2\hbar$k relative to the zero momentum peak, from the Fourier analysis of the TOF images \cite{suppmat}. For large $\as$, we observe the typical matter-wave interference pattern, as expected from a coherent state populating several lattice planes (see inset) \cite{Morsch2006dob}. As we lower $\as$, $C$ first remains fairly constant. For $\as$ below a certain critical value $\as^*\approx\SI{57}{\bohr}$, we observe a sudden loss of the interference pattern with a sharp decrease of $C$ to almost zero. 

Remarkably, we observe that this interaction-driven process is reversible. To test the restoring of the interference pattern, we employ the following protocol \cite{suppmat}: In brief, we first prepare the system in the lattice at constant and large $\as$ ($\as = \SI{69(2)}{\bohr}$). We then ramp down $\as$ below $\as^*$ ($\as = \SI{56(2)}{\bohr}$) in $\SI{20}{\milli\second}$ and wait until $C$ stabilizes to a small value; see Fig.\,\ref{Fig3}(b). Note that the interference pattern disappears after about $\SI{10}{\milli\second}$, which is on the order of the tunneling time $h/J$ between two neighboring lattice sites. 
At this point, we quench $\as$ back to its initial value and probe the time evolution of the system towards its new equilibrium state.
On a similar timescale, we observe the reappearance of the interference pattern with an increase of $C$, which then saturates to about $60\%$ of its initial value \footnote{The contrast is not fully recovered, which we attribute to the effect of inelastic losses.}. For comparison, we also show the data without inverting the field ramp.

The observed broad distribution in reciprocal space suggests that the system ground state has undergone a structural change, with the macroscopic wavefunction localized in one lattice plane. To verify this interpretation, we calculate the ground state of the system as a function of $\as$.
When the repulsive contact interaction dominates ($\as>a_\text{dd}$), we find an array of BECs occupying approximately three to five lattice planes; see insets Fig.\,\ref{Fig3}(a). In contrast, when the relative strength of the attractive dipolar interaction with respect to the other terms in the Hamiltonian is increased, the system reaches a critical point. Here, it undergoes a phase transition to a quasi-2D state, in which all atoms are localized into a single lattice plane to minimize their energy.
This purely interaction-driven phase transition--somewhat reminiscent of a continuous version of a superfluid to Mott insulator transition \cite{Greiner2002qpt}--is stabilized by quantum fluctuations (LHY), preventing the subsequent collapse of the system \cite{Maluckov2008sdn,Gligoric2008bsi}.
The predicted critical point occurs exactly where we observe the disappearance of the interference pattern in the experiments. We find an overall excellent agreement between the measured and the calculated $C$ from both the discrete 1D model and the 3D theory without any free fitting parameters, except for a rescaling factor to the contrast amplitude to account for the thermal atoms in the experiment.

\begin{figure}[t]
\centering
\includegraphics[width=\columnwidth]{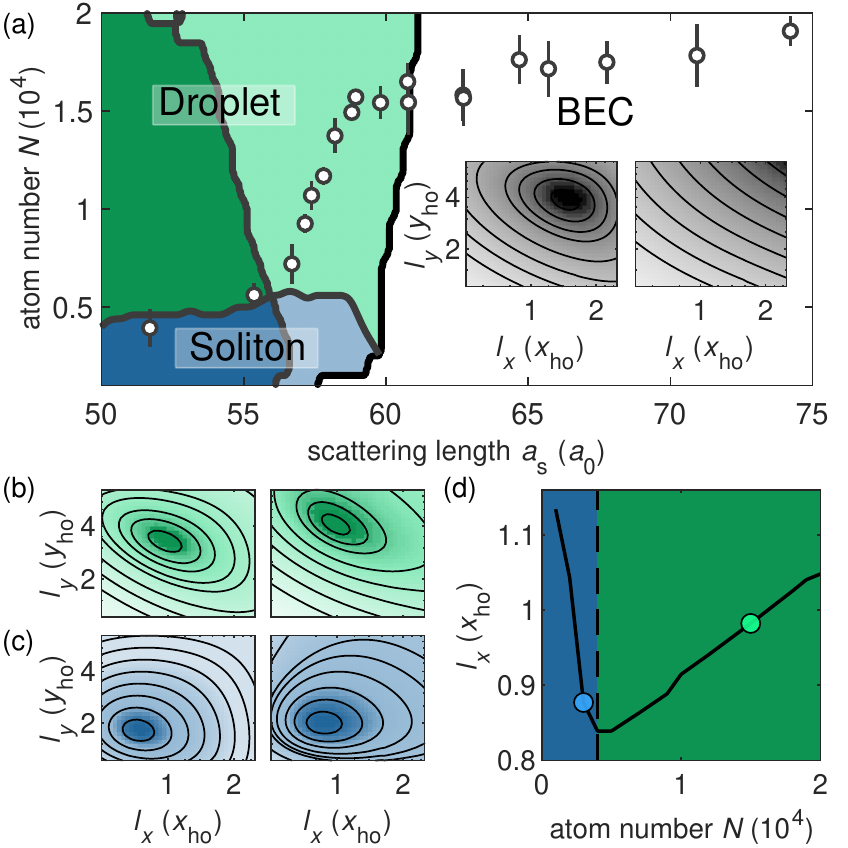}
\caption{\textbf{Phase diagram and energy landscapes.} \textbf{(a)}\,Phase diagram as a function of $\as$ and atom number. The white region denotes a trap-bound BEC extended over several lattice sites. The colored regions denote quasi-2D self-bound solutions: a droplet (green), a soliton (blue), each either extended over several lattice sites (lighter shade) or localized (darker shade, ${>}95\%$ of the atoms are localized in the central lattice plane). Circles show our experimental data points from Fig.\,\ref{Fig3}(a). \textbf{Inset (a), (b-c)}\,Energy landscapes as a function of the radial widths $l_x$ and $l_y$, in units of the radial harmonic oscillator lengths $x_\text{ho}=\SI{0.50(1)}{\micro\meter}$ and $y_\text{ho}=\SI{1.42(1)}{\micro\meter}$, respectively, with (left) and without (right) the radial harmonic trap, for (inset (a))\,BEC $(\as,\,N)= (\SI{70}{\bohr},\, 1.5\times10^4)$, (b)\,droplet $(\as,\,N)= (\SI{65}{\bohr},\, 1.5\times10^4)$ and (c)\,soliton $(\as,\,N)= (\SI{51.5}{\bohr},\,0.4\times10^4)$ regimes, with darker shading at the minima. \textbf{(d)}\,Radial width $l_x$ versus $N$ for $\as=\SI{51.5}{\bohr}$. The dashed line indicates the soliton-to-droplet transition point, and the circles indicate the position of (b-c).}
\label{Fig4}
\end{figure}

The observation of this phase transition to a quasi-2D localized state driven by interactions points to the existence of a rich variety of phases. The importance of the LHY correction and its peculiar density scaling motivate us to investigate the properties of the ground state as a function of $\as$ and atom number to identify distinct phases in this unique setting. For this, we employ our discrete model to derive a full phase diagram; see Fig.\,\ref{Fig4}(a). 
To investigate the boundness of the states, we assess the impact of the radial harmonic trap on the minimum of the variational energy, which is a function of the radial widths $l_x$ and $l_y$. 
At large scattering lengths, as expected, we find a stable delocalized BEC phase, where the total interaction energy (mean-field + LHY) is positive. The state is trap-bound, meaning that there is no energy minimum without the radial harmonic confinement; inset of Fig.\,\ref{Fig4}(a).

Reducing $\as$, we find an energy minimum even without the radial harmonic trap (colored region in the phase diagram). These quasi-2D self-bound solutions (the lattice still provides axial confinement) are either extended over several sites (lighter color) or localized to a single plane (darker color). In the literature, there are two paradigmatic examples of self-bound objects with attractive mean-field energy: droplets and solitons. Droplets can exist in one, two or three dimensions and are stabilized through the LHY correction \cite{chomaz2022dpa}. Stable bright solitons only exist in quasi-1D systems with attractive contact interactions and are stabilized against collapse purely by kinetic energy. In the search for solitons in higher dimensions, theoretical studies have suggested that the DDI could stabilize such 2D solutions \cite{Tikhonenkov2008asi,Raghunandan2015tdb}. To the best of our knowledge, there have been no studies on the effect the LHY correction has on this prediction, nor experimental observation. In the present case, where many interactions and kinetic energy compete, a classification of self-bound solutions is much less straightforward.
As a crucial distinction between a soliton and a droplet, we use the scaling of the system width with atom number. The soliton width (along the collapse direction) scales inversely with increasing atom number \cite{shabat1972exact}, while in contrast, the droplet size increases in all directions with $N$ \cite{pal2022infinite}, as predicted in a quasi-1D setting \cite{edmonds2020quantum}. We use this distinction to draw a boundary between the two phases, observing a phase transition at around 5000 atoms, for both single-site and multi-site solitons.
The overlaying of our measurements (Fig.\ref{Fig3}(a)) onto the phase diagram suggests that the experiments have already reached the interesting regimes of both 2D self-bound droplet and dipolar solitons. This opens the door to future experimental investigation on the self-bound nature and properties of these new 2D phases.

In conclusion, we theoretically and experimentally investigate the behavior of a strongly dipolar quantum gas in a 1D optical lattice. We employ BOs and characterize their dephasing rate as a function of $\as$. We observe a minimum in the dephasing shifted $\SI{6}{\bohr}$ away from the purely mean-field prediction, providing an interferometric measure of the beyond-mean-field contribution. 
For low enough $\as$, the system enters into a quasi-2D state which is localized onto a single lattice plane, providing a genuine interaction-driven path to reach reduced dimensions in dipolar gases.
Using our developed discrete theory model, we derive a full phase diagram which confirms the observed localization transition.
This also reveals signatures of quasi-2D self-bound dipolar droplet solutions, and the long sought-after 2D anisotropic dipolar soliton, first predicted in Ref.\,\cite{Tikhonenkov2008asi} (see also \cite{Pedri2005tdb,Raghunandan2015tdb}).
Our work paves the way for future studies of the soliton-to-droplet crossover in a dipolar gas, as observed in a Bose-Bose gas \cite{Cheiney2018bst}, and of the ``solitonic" nature \cite{drazin1989solitons} of dipolar solitary waves \cite{cuevas2009solitons,eichler2012collisions,adhikari2014bright,baizakov2015interaction,edmonds2017engineering}.

\begin{acknowledgments}
We thank R.\,N.\,Bisset, A.\,Houwman, L.\,Lavoine, R.\,Grimm, and L.\,Tarruell for stimulating discussions, and B.\,Yang for his support in the early stage of the experiment. This work is financially supported through an ERC Consolidator Grant (RARE, no.\,681432) and a DFG/FWF (FOR 2247/I4317-N36). We also acknowledge the Innsbruck Laser Core Facility, financed by the Austrian Federal Ministry of Science, Research and Economy. Part of the computational results presented have been achieved using the HPC infrastructure LEO of the University of Innsbruck.
\end{acknowledgments}

* Correspondence and requests for materials should be addressed to Francesca.Ferlaino@uibk.ac.at.

\bibliographystyle{apsrev4-2}
\bibliography{dipolar_latest, BO_bib}

\clearpage
\appendix
\onecolumngrid
\begin{center}
    {\bf\large Supplemental materials: Strongly dipolar gases in a one-dimensional lattice: Bloch oscillations and matter-wave localization}\\\vspace{0.3cm}
    {\normalsize G. Natale, T. Bland, S. Gschwendtner, L. Lafforgue, D. S. Gr\"un, A. Patscheider, M. J. Mark, and F. Ferlaino}
\end{center}
\hspace{3cm}
\twocolumngrid
\makeatletter
\renewcommand{\theequation}{S\arabic{equation}}
\renewcommand{\thefigure}{S\arabic{figure}}
\setcounter{equation}{0}
\setcounter{figure}{0}

\section*{Theoretical model}
In this work, we use an extended Gross-Pitaevskii theory for direct comparison to our experimental results. We employ both the standard three-dimensional form of the extended Gross-Pitaevskii equation (eGPE) and derive a discrete effective one-dimensional eGPE. Starting with the three-dimensional case, our system can be described by the 3D eGPE of the form \cite{Wachtler2016qfi,Bisset2016gsp,FerrierBarbut2016ooq,Chomaz2016qfd}
\begin{align}
    i \hbar \diffp{}{t} \Psi(\vec{x}, t) = \Big{[} &- \displaystyle\frac{\hbar^2}{2 m} \nabla^2 + V_\mathrm{harm}(\vec{x}) \nonumber\\
    &+ V_\mathrm{latt}(z) - F_\text{ext} z + g |\Psi(\vec{x}, t)|^2 \nonumber\\
    &+ \int \text{d}^3\vec{x}' \, U_\text{dd}(\vec{x} - \vec{x}')|\Psi(\vec{x}', t)|^2\nonumber\\
    &+ \gamma_\text{QF} |\Psi(\vec{x}, t)|^{3} \Big{]} \Psi(\vec{x}, t)\,,
    \label{eqn:eGPE}
\end{align}
where the wavefunction $\Psi$ is normalized to the total atom number $N=\int {\rm d}^3\vec{x}\,|\Psi|^2$. The atoms are confined in a harmonicpotential $V_\text{harm} = \sum_{\xi = x,y,z} \frac{1}{2} m \omega_\xi^2 \xi^2$ with single particle mass $m$ and trap frequencies $\omega_\xi$, together with the lattice potential $V_\text{latt} = s E_\text{rec} \sin^2{(k z)}$ where $s$ is the tunable lattice depth in multiples of the recoil energy $E_\text{rec}$ and $k=2\pi/\lambda$ is the lattice spacing in reciprocal space. The mean-field interaction contributions are $g = 4 \pi \hbar^2 a_\text{s}/m$ for the contact interaction, governed by the s-wave scattering length $a_\text{s}$, and the long-ranged anisotropic dipolar interaction potential $U_\text{dd}(\vec{x}) = 3\hbar^2a_\text{dd}/m\left(1-3\cos^2\theta\right)/|\vec{x}|^3$, where $a_\text{dd} = \mu_0\mu_m^2m/12\pi\hbar^2$ with magnetic moment $\mu_m$ and $\theta$ is the angle between the polarization axis ($y$-axis) and the vector between neighboring atoms. We also include beyond-mean-field effects through the quantum fluctuations term $\gamma_\text{QF} = \frac{32}{3} g \sqrt{\frac{a^3_\mathrm{s}}{\pi}} \left( 1 + \frac{3}{2}\varepsilon^2_\mathrm{dd} \right)$ \cite{Lima2011qfi}, which depends on the relative strength between the dipolar and short-ranged interactions $\varepsilon_\text{dd} = a_\text{dd}/a_\text{s}$. Finally, $F_\text{ext} = g_\text{grav} m$ denotes the external force exerted on the system by gravity.

In this work, we employ the imaginary time-evolution technique on Eq.\,\eqref{eqn:eGPE} in order to find stationary solutions for the wavefunction in the lattice, without gravity. For various atom numbers and scattering lengths, we use a numerical grid of lengths $(L_x,\,L_y,\,L_z) = (6,\,33.3,\,6)\,\mu$m, with corresponding grid points $128\times256\times128$. The dipolar term is efficiently calculated in momentum space, and we use a cylindrical cut-off in order to negate the effects of aliasing from the Fourier transforms \cite{Lu2010sdo}.

To derive the effective one-dimensional model, we follow Ref.\,\cite{Blakie2020vtf} by assuming a wavefunction decomposition
\begin{align}
    {\Psi(\vec{x}, t) = \Phi(x,y,l,\eta)\psi(z,t)\equiv\frac{1}{\sqrt{\pi} l} e^{- (\eta x^2 + y^2/\eta)/2 l^2} \psi(z, t)}\,,
    \label{eqn:variational}    
\end{align} 
with variational parameters $l$ and $\eta$ representing the width of the radial wavefunction and the anisotropy of the state, respectively. Integrating out the transverse directions $(x,y)$ in Eq.\,\eqref{eqn:eGPE} upon substitution of the ansatz above gives the continuous quasi-one-dimensional eGPE, which when combined with a variational minimization of the energy functional to find $(l,\eta)$ gives close agreement to the full 3D eGPE \cite{Blakie2020vtf}. We further decompose the longitudinal wave function $\psi(z, t)$ into a sum of Wannier functions $w(z)$ of the lowest energy band over all lattice sites
\begin{align}
    \psi(z, t) = \sqrt{N} \sum_j c_j(t)\,w(z - z_j)\,,
    \label{eqn:wann}
\end{align}
for complex amplitudes $c_j$, and positions of lattice minima $z_j = (\lambda/2)j$. For deep enough lattices, the Wannier functions are well approximated by Gaussians of the form $w(z) = \left(\pi l_\text{latt}^2\right)^{-1/4}e^{-z^2/2 l_\mathrm{latt}^2}$, with $l_\mathrm{latt} = (k \sqrt[4]{s})^{-1}$. After multiplying on the left by $c_j^*$ and integrating over $z$, we obtain a set of discrete effective one-dimensional eGPEs
\begin{widetext}
\begin{equation}
\label{eq:Discrete_model}
    i \hbar \diffp{c_j}{t} = - J(c_{j+1} + c_{j-1}) + \left(-F_\mathrm{ext}z_j + V_\text{harm}(z) + g^\text{1D}N|c_j|^2 + N\sum_{k} U_{|j-k|}^\mathrm{dd}|c_k|^2 + \gamma^\mathrm{1D}_\mathrm{QF}N^{3/2} \gamma_\mathrm{QF} |c_j|^{3} \right) c_j \,,
\end{equation}
\end{widetext}
with the reduced effective one-dimensional parameters $\gamma^\mathrm{1D}_\mathrm{QF} = 2^{3/2} / (5 \pi^{3/2} l^2 l_\mathrm{latt})^{3/2} \gamma_\mathrm{QF}$ and $g^\text{1D} = g / ((2 \pi)^{3/2} l^2 l_\mathrm{latt})$. 
Here, $J$ denotes the tunneling rate between two neighboring lattice sites.
The dipolar interaction coefficients between lattice sites $j$ and $k$ depend both on the separation $|j-k|$, and non-trivially on the size $l$ and anisotropy $\eta$ of the transverse cloud. For the variational minimization, we generate an interpolating function for a sensible range of $(l,\eta)$ and separations up to $|j-k|=6$ via
\begin{align}
    U_{|j-k|}^\mathrm{dd}(l,\eta) = \int \text{d}^3\vec{x}&\Big\{ |\Psi_0(\vec{x}-z_{|j-k|}\hat{e}_z,l,\eta)|^2 \nonumber \\ &\int\text{d}^3\vec{x}'\, U_\text{dd}(\vec{x} - \vec{x}')|\Psi_0(\vec{x}',l,\eta)|^2\Big\}\,,
\end{align}
where $\Psi_0(\vec{x}',l,\eta) = \Phi(x,y,l,\eta)w(z)$ [see Eqs.\,\eqref{eqn:variational} and \eqref{eqn:wann}]. This allows us to simply look up the values of $U_{|j-k|}^\mathrm{dd}$ without having to recalculate for every time step during the energy minimization. We note that the energy contribution rapidly declines for separations larger than 2 sites, and find that 6 is more than sufficient to quantitatively describe the physics.

To find the stationary state solution of Eq.\,\eqref{eq:Discrete_model} (without gravity) we employ an imaginary time-evolution in combination with an optimization scheme, aiming to find the state which minimizes the total energy functional 
\begin{equation}
\label{eq:Total_energy_functional}
    \mathcal{E}[\textbf{c}; l, \eta] = \mathcal{E}_\perp [l, \eta] + \mathcal{E}_\parallel [\textbf{c}; l, \eta] \ ,
\end{equation}
where $\textbf{c}=(c_1,c_2,\dots,c_n)$ for $n$ total lattice sites.
Here, $\mathcal{E}_\perp [l, \eta]$ gives the energy contribution from the transverse variational wave function, which reads 
\begin{equation}
    \mathcal{E}_\perp [l, \eta] = \frac{\hbar^2}{2 m l^2} \left(\eta + \frac{1}{\eta}\right) + \frac{m l^2}{4} \left( \frac{\omega^2_x}{\eta} + \eta \omega^2_y \right) \ .
\end{equation}
The latter term of Eq.\,\eqref{eq:Total_energy_functional} gives the discrete energy functional for the amplitudes $c_j$, which includes the tunneling and all interaction terms
\begin{align}
    \mathcal{E}_\parallel [\textbf{c}; l, \eta] = & - \sum_j J (c_{j+1} + c_{j-1}) c_j \nonumber \\
     & + \frac{1}{2} N g^\text{1D} \sum_j |c_j|^4 + \frac{1}{2}N \sum_{j,k} U^\text{dd}_{|j-k|} |c_k|^2 |c_j|^2 \nonumber \\
    %  & + \frac{1}{2} \sum_j \left[ N g^\text{1D} |c_j|^4 + \sum_{k} U^\text{dd}_{|j-k|} |c_k|^2 |c_j|^2 \right] \\ % doesn't save space
     & + \frac{2}{5} N^{3/2} \gamma^\text{1D}_\mathrm{QF} \sum_j |c_j|^5\,. \label{eq:Axial_energy_functional}
\end{align}

Starting from an initial distribution of the amplitudes $c_j$ we first determine the variational parameters $(l,\,\eta)$, which is done via an optimization scheme minimizing Eq.\,\eqref{eq:Axial_energy_functional}. Subsequently, we evolve the amplitudes in imaginary time using Eq.\,\eqref{eq:Discrete_model} and repeat this process until we find the minimum of the total energy function Eq.\,\eqref{eq:Total_energy_functional}. 

\begin{figure}[t]
\centering
\includegraphics[width=\columnwidth]{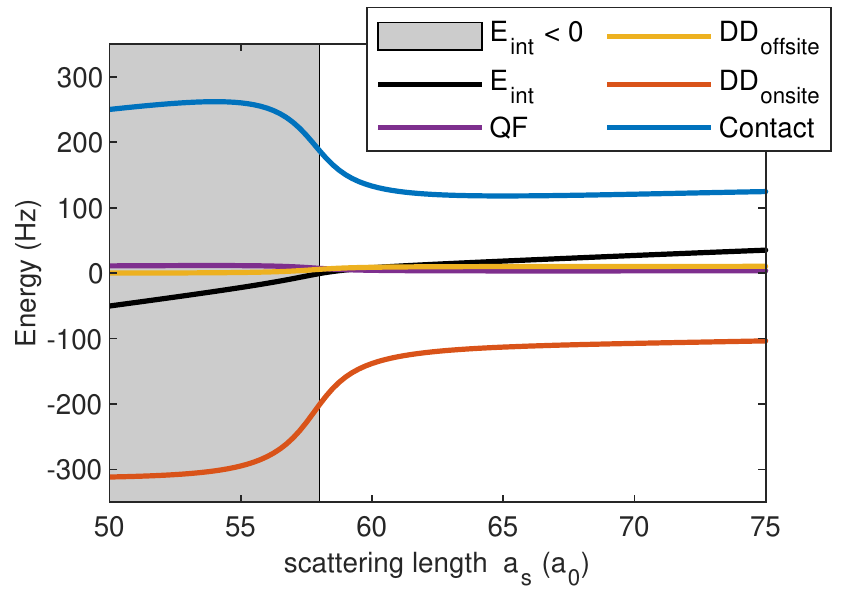} %0.75\textwidth
\caption{\label{figS1}\textbf{Interaction energy contributions.} Scattering length dependency of the individual interaction contributions of the ground state solutions from the 1D model, calculated for $N=10^4$ atoms.}
\end{figure}

In Fig.\,\ref{figS1} we assess the different interaction energy contributions to Eq.\,\eqref{eq:Total_energy_functional} for a range of scattering lengths. For $\as>a_{\rm dd}$ the total interaction energy is positive, and it corresponds to a dilute BEC. Following $\as$ to smaller values all interaction contributions are almost constant, until at around $\as=\SI{60}{\bohr}$ there is a phase transition from the BEC to droplet state, as identified in Fig.\,4 of the main text. This sharp gradient ceases at around $\as=\SI{55}{\bohr}$, where the atoms are localized to a single lattice plane. Note that although the DDI offsite energy is typically only 10\% of the onsite counterpart, it constitutes a significant contribution to the total interaction energy in the system, shifting the BEC to droplet crossover and localization transitions by a few \SI{}{\bohr}.

Once we have the ground state of the system, we employ the discrete effective one-dimensional eGPE in real-time to simulate the Bloch oscillations in the presence of gravity. 

\section*{2D to 3D crossover}
The dimensionality of the system is known to highly influence the size and even the sign of the beyond-mean-field contribution, in both Bose-Bose \cite{Petrov2016uld,ilg2018dimensional,Lavoine2021bmf} and dipolar \cite{Edler2017qfi,boudjemaa2019two,zin2021quantum} gases. Here, we assess the validity of employing the full 3D LHY correction to our system. Following Ref.\,\cite{zin2021quantum}, we define the dimensionless parameter $\xi = g n /\epsilon_0$--dependent on the contact interactions $g$, peak 3D density $n$, and the confinement energy scale $\epsilon_0 = \hbar^2\pi^2/2mz_\text{ho}^2$--that indicates which dimensionality regime our system is in. If $\xi\gtrsim1$ we are safe to use the 3D LHY term, whereas if $\xi\ll1$ the 2D solution deviates from the 3D one. Deep in the localized droplet regime, where the peak density is on the order of $10^{22}\SI{}{\per\cubic\meter}$, we find $\xi\approx2$, and the 3D LHY as used throughout this work is valid. Even at large scattering lengths, where the peak density is closer to $5\times10^{20}\SI{}{\per\cubic\meter}$, we find $\xi\approx0.5$, which introduces an error of less than 5\% between the 2D and 3D LHY terms \cite{zin2021quantum}. In this limit, the 2D LHY term may be more appropriate, however in the dilute BEC phase the impact of the LHY is minimal.

\section*{Analytic model of dephasing}
 
Starting from the discrete 1D eGPE we decompose the coefficients $c_j$ into amplitude and phase as $c_j = |c_j|\exp(-i\phi_j)$, and then integrate Eq.\,\eqref{eq:Discrete_model} in time to give
\begin{align}
\phi_j(t) &= \Bigg(- F_\mathrm{ext} z_j +g^\text{1D}N|c_j|^2 + N\sum_{k} U_{|j-k|}^\mathrm{dd}|c_k|^2 \nonumber\\ &\qquad\qquad\qquad\qquad+ \gamma^\mathrm{1D}_\mathrm{QF}N^{3/2} \gamma_\mathrm{QF} |c_j|^{3}  \Bigg) \frac{t}{\hbar} \nonumber\\ &\equiv \left(- F_\mathrm{ext} z_j + \mu_j\right)\frac{t}{\hbar}\,,
\end{align}
with onsite chemical potentials $\mu_j$, and where we have also assumed that $F_\mathrm{ext}d\gg J$ such that the amplitudes $|c_j|$ are frozen.

\begin{figure}
    \centering
    \includegraphics[width=\columnwidth]{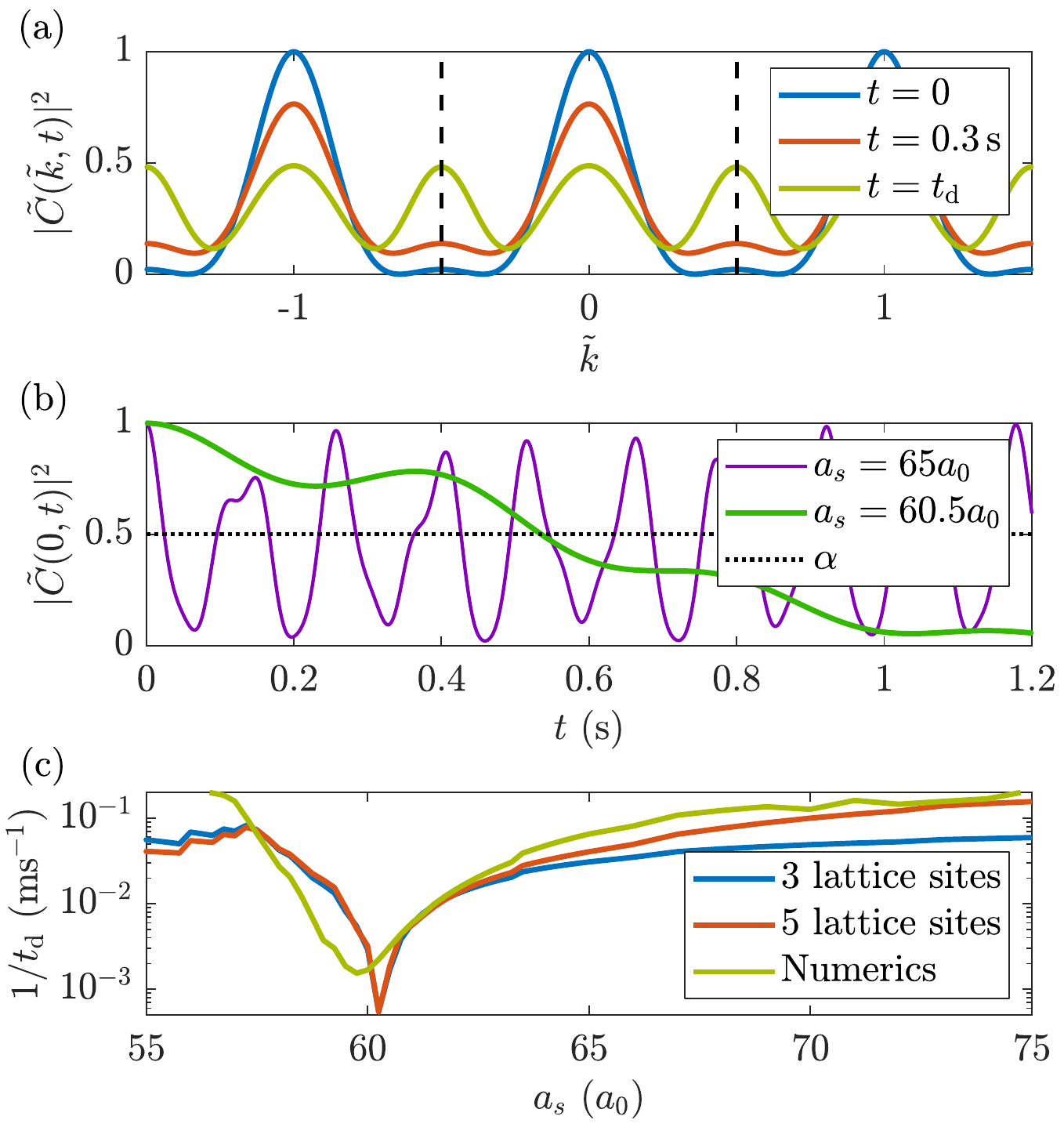}
    \caption{\textbf{Analytic dephasing rate. (a)}\,Evolution of the function $\tilde{C}$, with $\tilde{k}$ normalized to the Brillouin zone in the moving frame, and $\as = 60.5\,a_0$. Here, the solution of Eq.\,\eqref{eqn:t_d} is $t_{\rm d} = 0.59$s. \textbf{(b)}\,Time evolution of the central point of $\tilde{C}$, showing when  $|\tilde{C}|^2$ crosses $\alpha=0.5$. The function $\tilde{C}$ is scaled to the value at $\tilde{C}(0,0)$. \textbf{(c)}\,Analytic dephasing rate ($\gamma = 1/t_{\rm d}$) obtained for the 3 lattice site approximation Eq.\,\eqref{eqn:t_d} and the 5 lattice site approximation Eq.\,\eqref{eqn:t_d5}, compared to the numerically obtained value from a real-time simulation of the discrete model, Eq.\,\eqref{eq:Discrete_model}.}
    \label{fig:deph_thry}
\end{figure}

Following Ref.\,\cite{Witthaut2005boo}, we write the Fourier transform of the quasi-1D wavefunction as
\begin{align}
    \psi(k,t) = w(k)\sum_j |c_j|\exp[-i(kz_j + \phi_j(t))] = w(k) \tilde{C}(k,t) \,,
    \end{align}
where $w(k)$ is the momentum space Wannier function, and phases $\phi_j$ are given above. If all interactions are set to zero this function is initially a delta function situated at $k=0$ and moves in $k$-space as $\tilde{k} = k-F_\text{ext}t/\hbar$. Interactions broaden $\tilde{C}(k,t)$, leading to a dephasing of coefficients $c_j$.
Fig.\,\ref{fig:deph_thry}(a) depicts $|\tilde{C}(k,t)|^2$ as a function of $k$ at different times $t$, normalized to $|\tilde{C}(0,0)|^2$.

We extract an analytic approximation to the dephasing time by considering the temporal behaviour of the point $|\tilde{C}(0,t)|^2$, i.e.\,at $\tilde{k}=0$. During dephasing this point rapidly decreases through interference between neighboring sites. This quantity is plotted in Fig.\,\ref{fig:deph_thry}(b) for a few example scattering lengths. It reaches the threshold $\alpha / \tilde{C}(0,0)= 0.5$ at the dephasing time $t = t_{\rm d}$, where many $k$-modes are now highly occupied. This time can be found through the smallest positive solution of
\begin{align}
    \alpha = \left|\left(\sum_j|c_j|\cos\left(\frac{\mu_jt_{\rm d}}{\hbar}\right)\right)^2 +\left(\sum_j|c_j|\sin\left(\frac{\mu_jt_{\rm d}}{\hbar}\right)\right)^2\right|\,.
    \label{eqn:alph}
\end{align}
Exact solutions to $|\tilde{C}(0,t_{\rm d})|^2=\alpha$ can be only found in limiting cases. For the three lattice site case, with $j=-1,0,1$ and noting the symmetry of $|c_j| = |c_{-j}|$ we obtain
\begin{align}
    t_{\rm d} = \left|\arccos\left(\frac{4|c_0c_1|}{\alpha + |c_0|^2-2}\right)\frac{\hbar}{(\mu_1-\mu_0)}\right|\,,
    \label{eqn:t_d}
\end{align}
This relation is expected to give an accurate prediction of the dephasing time for all states where only 3 lattice sites are dominant. From this equation, one can see how the dephasing time tends to infinity in the limit of equally distributed chemical potentials, as observed in Fig.\,2 of the main text. We can extend this to 5 sites, but it is not as trivial. One needs to numerically solve the transcendental equation
\begin{align}
    \alpha = &\Big|2 - |c_0|^2 + 4|c_0c_1|\cos\left(\frac{(\mu_0-\mu_1)t_{\rm d}}{\hbar}\right) \nonumber\\ &+ 4|c_0c_2|\cos\left(\frac{(\mu_0-\mu_2)t_{\rm d}}{\hbar}\right) \nonumber\\ &+ 8|c_1c_2|\cos\left(\frac{(\mu_1-\mu_2)t_{\rm d}}{\hbar}\right)\Big|\,,
    \label{eqn:t_d5}
\end{align}
for the smallest non-zero root $t_{\rm d}$. We compare the results from Eqs.\,\eqref{eqn:t_d} and \eqref{eqn:t_d5} to the numerically obtained dephasing rate, $\gamma=1/t_{\rm d}$, in Fig.\,\ref{fig:deph_thry}(c), as presented in Fig.\,2 of the main text, and find excellent agreement.

\section*{Experimental protocol}
We prepare a $\Er$ spin-polarized BEC similar to Ref.\,\cite{Chomaz2016qfd}. The magnetic field during the evaporation is along the z-axis with absolute value $\mathbf{|B|}=B_z= \SI{1.9}{\gauss}$ ($\as=\SI{80(1)}{\bohr})$, see Fig.\,\ref{Fig1}(a). The B-to-as conversion has been precisely mapped out in previous experiments \cite{Chomaz2016qfd,patscheider2021accurate}. Before loading the lattice, we rotate the magnetic field direction along the y-axis in \SI{50}{ms} and change its absolute value to set the scattering length.
At this step, we typically achieve $5 \times 10^4$ atoms with more than 60 \% condensed fraction in a cigar shape dipole trap with trapping frequencies $\omega_{x,y,z} = 2 \pi \,(240(3),30(3),217(1))$\,\SI{}{\hertz}. For our experiments, the atoms are then loaded in a 1D lattice by a \SI{20}{\milli\second} exponential ramp of the lattice depth. 
This is the experimental protocol used in Fig\,\ref{Fig1}, \ref{Fig2}, and \ref{Fig3}(a).

To study the reversibility of the interaction-induced transition to a single lattice site (\ref{Fig3}(b)), i.e. the evolution of the contrast due to a change of the scattering length, we employ a different protocol from the one above. In fact, in our experiment, the magnetic field along the y-direction can be changed on a timescale of $\simeq \SI{20}{\milli\second}$, which is slower compared to the z-direction ($\simeq \SI{1}{\milli\second}$). For this dataset, we prepare the BEC with $\mathbf{B}=(0,0.25,1)\,\SI{}{\gauss}$ and then we load the lattice as described above. We then linearly ramp the field in \SI{20}{\milli\second} to $\mathbf{B}=(0,0.25,0)\,\SI{}{\gauss}$ and record the time evolution. In Fig. \ref{Fig3}(b), we study the contrast evolution after the ramp. 
For the black dataset, the magnetic field is quenched back to the initial value after \SI{10}{\milli\second}.

For Fig\,\ref{Fig4}, we extract the atom number condensed in the lattice by releasing the cloud from the combined ODT-lattice trap and by performing an absorption imaging after \SI{30}{\milli\second} of TOF.  We integrate the density along the lattice axis and use a double Gaussian fit on the integrated density profile. We repeat the sequence 4-8 times for every scattering length. At low scattering lengths, we find a decreased number of condensed atoms, see Fig.\ref{Fig4}.  We attribute this to an increase of three-body loss in the vicinity of a Feshbach resonance \cite{Chomaz2016qfd} and the increased density of the groundstate.

\section*{Analysis of momentum distribution during Bloch Oscillation }
\begin{figure}[t]
\includegraphics[width=\columnwidth]{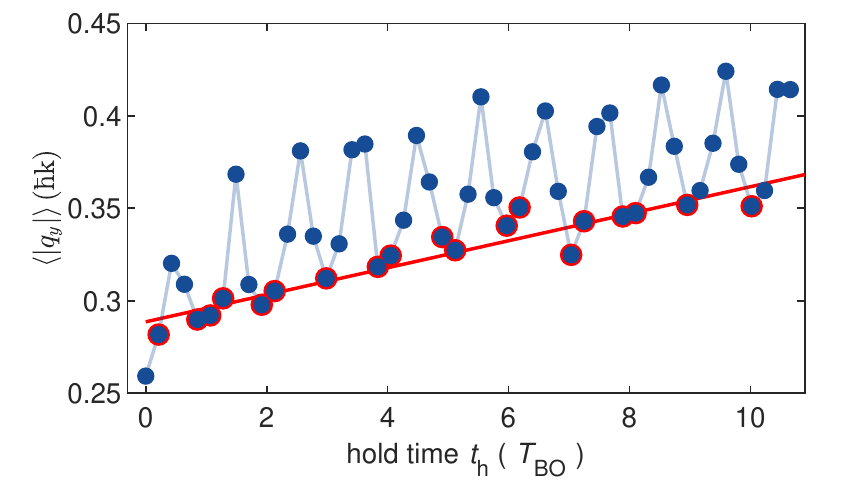} %0.75\textwidth
\caption{\label{figS1meank}\textbf{Evolution of the \meank. } In the figure \meank as a function of time for \as=\SI{58.8}{\bohr}. The points with red edges are the one selected for the fit. The red line corresponds to the fit result $A(t-\tau)+0.5$}
\end{figure}

When the Bloch oscillation dephases, the width of the momentum distribution increases with time \cite{Morsch2001boa}. To evaluate the dephasing rate we analyze the 1D momentum distribution along z, $n(q_z)$,  as a function of the holding time. %$n(q_z)$ is obtained after numerical integration along $q_x$  and normalization.
Because of our limited vertical optical access, the 1D lattice is not perfectly aligned with the z (gravity) direction.
We measure a tilt of \SI{9(1)}{\degree}. Such a tilt effectively weakens the radial trapping strength, limiting our observation time to \SI{12}{\milli\second}, which anyhow allows us to observe up to 25 BO period. 

From $n(q_z)$, we can extract the maximum position ($q_z^{\text{max}}$) and the quantity \meank, given by \[\meank = \sum_{q_z} n(q_z)\left| q_z -q_z^{\text{max}}\right|.\] This quantity is proportional to the width of the distribution.
In Fig.\,\ref{figS1meank}, we report \meank for \as = 65.7(1.0)\,\SI{}{\bohr}. 
\begin{figure}[t]
\centering
\includegraphics[width=\columnwidth]{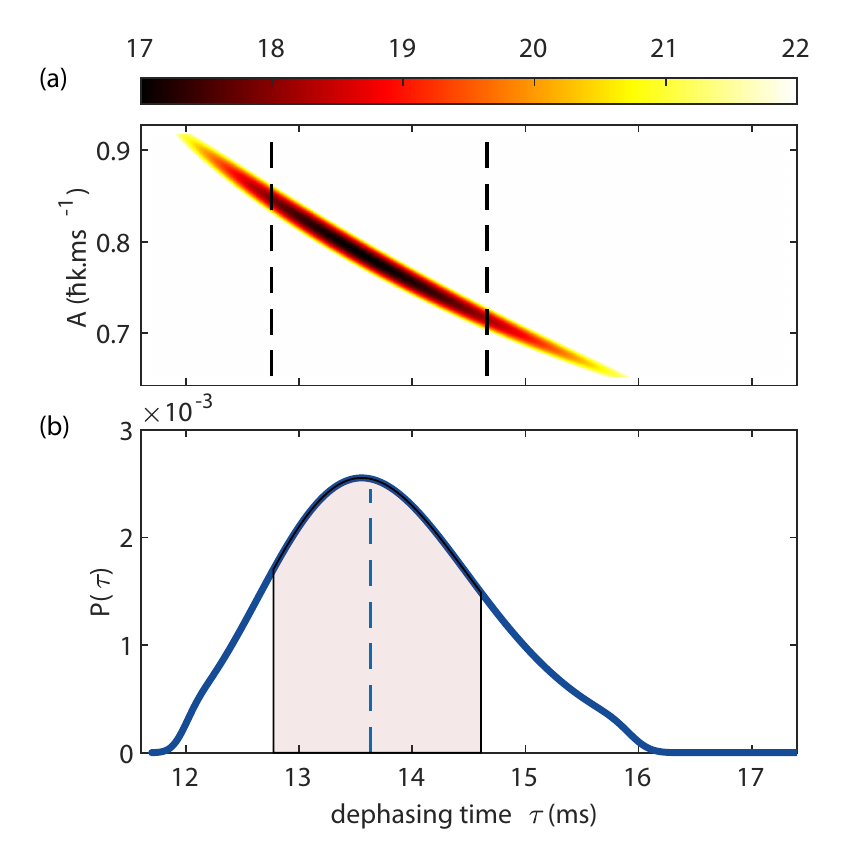} 
\caption{\label{figS2} \textbf{Uncertainties $\chi^2$ analysis}. \textbf{(a)}\,$\chi^2$ as a function of A and $\tau$. The black dash lines correspond to the value $\chi_{min}^2+1$. Below, \textbf{(b)}\,probability distribution of $\tau$, given by numerical integration of $\int_A dA e^{-\chi^2(A,\tau)/2}$ and normalization to 1. \as=\SI{58.8}{\bohr}. The dashed line indicates the fit result and the shaded area the 68\% confidence interval.}
\end{figure}
To quantify the dephasing rate $\gamma$, we apply a linear fit to \meank. For the fit, we select only the points at the center of the Brillouin zone, up to the time when \meank is reaching the fully dephased configuration, $0.5 \hbar k$. Indeed, when the cloud is at the edge of the Brillouin zone, \meank is artificially increased and it does not represent the dephasing, as shown in Fig.\,\ref{figS1meank}. We define the dephasing rate $\gamma$ as the inverse of the time $\tau$ that the fitted function needs to reach the value $0.5\, \hbar k$. Thus, using the fit parametrization $A(t-\tau)+0.5$, where $A$ and $\tau$ are the fitting variables and $t$ is the time, we can directly extract $\tau$ and its inverse $\gamma$. 

To determine the uncertainties with our non-linear parametrization, we analyze the $\chi^2(A,\tau)$.
We estimate the uncertainties on our data points by assuming equal statistical fluctuations around our fitting model and using the expected value $\left< \chi^2 \right>$= $N_{\text{data}}-2$. Figure\,\ref{figS2} shows a clear asymmetric shape for $\chi^2$, indicating asymmetric uncertainties on our fit parameters. As we are only interested in the uncertainties on $\tau$, we consider $P_{\chi^2}(\tau)=\frac{1}{N}\int_A dA e^{-\chi^2(A,\tau)/2}$, with $N$ a normalization constant. $P_{\chi^2}(\tau)$ corresponds to the probability distribution of $\tau$ for our fitting model. Finally, from $P_{\chi^2}(\tau)$, we define the 68\% confident interval of our dephasing rate $\gamma$ shown in Fig.\,\ref{figS2}.

In order to compare our experimental data with the theoretical predictions, we repeat the same analysis with the data from the 1D discrete model. 
Since in the experiment the condensed atom number changes with the scattering length, see Fig.\,\ref{Fig4}, the atom number considered in the theoretical simulations varies accordingly. 
In Fig.\,\ref{Fig2}, we account for the experimental fluctuations by taking an interval of $\pm20\%$ of the BEC atoms number. For each scattering length, we determine the extreme values of $\gamma$ in the $\pm20\%$ range, which we use to create the shaded area. 

 \section*{Contrast of the interference pattern}
The density modulation that usually characterizes a BEC loaded into a 1D lattice can be experimentally extracted from the matter-wave interferometry after a TOF expansion \cite{Greiner2002qpt}. To study the transition to one single occupied lattice site, we record the density distribution as a function of \as. 
In more details, for each picture we perform a Fourier transform (FT) of the integrated momentum distribution, $n(q_z)$. In the contact dominated regime, the lattice induces two sidepeaks at $\pm q_z^*$ in $n(q_z)$. Consequently, in the FT analysis the peaks are at $z^{*} \simeq \lambda_{\text{lattice}}$. The visibility of the interference pattern is then estimated as $n_{\text{FT}}(|z^*|)/n_{\text{FT}}(0)$.

\end{document}